\documentclass[3p,times,procedia]{elsarticle}
\usepackage[latin9]{inputenc}
\usepackage{amstext}
\usepackage{amssymb}
\usepackage{esint}

\makeatletter

\usepackage{amstext}
\usepackage{nupha_ecrc}

\volume{00}

\firstpage{1}

\journalname{Nuclear Physics A}

\runauth{}

\jid{nupha}

\jnltitlelogo{Nuclear Physics A}

\usepackage{amssymb}

\usepackage[figuresright]{rotating}

\begin{document}

\begin{frontmatter}



\dochead{XXVIIIth International Conference on Ultrarelativistic Nucleus-Nucleus Collisions\\ (Quark Matter 2019)}

\title{Quark polarization from parton scatterings in heavy ion collisions}


\author{Jun-Jie Zhang$^1$, Ren-Hong Fang$^2$, Qun Wang$^1$, Xin-Nian Wang$^3$}
\address{$^1$ Department of Modern Physics, University of Science and Technology of China, Hefei, Anhui 230026, China\\
$^2$ Key  Laboratory  of  Particle  Physics  and  Particle  Irradiation  (MOE),
Institute  of  Frontier  and  Interdisciplinary  Science, Shandong  University,  Qingdao,  Shandong  266237,  China\\
$^3$ Nuclear Science Division, MS 70R0319, Lawrence Berkeley National Laboratory, Berkeley, California 94720, USA
}

\begin{abstract}
Based on the result obtained in Ref. \cite{Zhang:2019xya},
we calculate the polarization of quarks from parton
scatterings in high energy heavy ion collisions.
The result is compared with the STAR data at $\sqrt{s_{NN}}=200$ GeV.
\end{abstract}


\end{frontmatter}



\section{Introduction}
A very large orbital angular momentum (OAM) can be generated in peripheral heavy ion collisions
in the direction perpendicular to the reaction plane. Such a large OAM in the system
can be converted into the spin polarization of hadrons along the direction of the OAM through
spin-orbit couplings \cite{Liang:2004ph}, see, e.g., Ref. \cite{Wang:2017jpl} for a recent review.
This effect is called the global polarization and differs from the polarization effect
of a particle with respect to the production plane which depends on the particle's momentum.
Non-vanishing global polarization of $\Lambda$ and $\overline{\Lambda}$ hyperons has been measured by
the STAR collaboration \cite{STAR:2017ckg,Adam:2018ivw}.
Various theoretical models have been proposed to describe the global polarization effect
\cite{Liang:2004ph,Liang:2004xn,Voloshin:2004ha,Betz:2007kg,Becattini:2007sr,Gao:2007bc,Huang:2011ru,Becattini:2013fla,
Florkowski:2017ruc,Fang:2016vpj,Weickgenannt:2019dks}.
Recently we proposed a microscopic model for the global polarization from particle scatterings \cite{Zhang:2019xya}.
The model does not rely on the assumption that the spin degree of freedom has reached a local equilibrium.
The spin-vorticity coupling naturally emerges from scatterings of particles at different
space-time points which incorporate polarized scattering amplitudes with the spin-orbit coupling \cite{Gao:2007bc}.

\section{Polarization rate for spin-1/2 particles}
In this note, we apply the result of Ref. \cite{Zhang:2019xya} to calculate the global polarization
in heavy ion collisions and compare it with the STAR data at $\sqrt{s_{NN}}=200$ GeV \cite{Adam:2018ivw}.
We consider all 2-to-2 parton scattering processes $A+B\rightarrow 1+2$ with at least one quark or anti-quark
in the final state. Here $A$ and $B$ represent incoming partons, and 1 and 2 represent outgoing partons
in which 2 is chosen to be the quark or anti-quark.
In this section, we will denote a quantity in the center of mass (CMS) frame with a subscript 'c'.
There is no 'c' index for a quantity in the laboratory frame.
The spin asymmetry rate for the quark (as parton 2) per unit volume at the space-time point $X$
is given by Eqs. (21,22) of Ref. \cite{Zhang:2019xya},
\begin{eqnarray}
\frac{d^{4}\mathbf{P}_{AB\rightarrow 1q}(X)}{dX^{4}} & = & -\frac{1}{(2\pi)^{4}}\int\frac{d^{3}p_{A}}{(2\pi)^{3}2E_{A}}\frac{d^{3}p_{B}}{(2\pi)^{3}2E_{B}}\frac{d^{3}p_{c,1}}{(2\pi)^{3}2E_{c,1}}\frac{d^{3}p_{c,2}}{(2\pi)^{3}2E_{c,2}}\nonumber \\
 &  & \times|v_{c,A}-v_{c,B}|\int d^{3}k_{c,A}d^{3}k_{c,B}d^{3}k_{c,A}^{\prime}d^{3}k_{c,B}^{\prime}\nonumber \\
 &  & \times\phi_{A}(\mathbf{k}_{c,A}-\mathbf{p}_{c,A})\phi_{B}(\mathbf{k}_{c,B}-\mathbf{p}_{c,B})\phi_{A}^{*}(\mathbf{k}_{c,A}^{\prime}-\mathbf{p}_{c,A})\phi_{B}^{*}(\mathbf{k}_{c,B}^{\prime}-\mathbf{p}_{c,B})\nonumber \\
 &  & \times\delta^{(4)}(k_{c,A}^{\prime}+k_{c,B}^{\prime}-p_{c,1}-p_{c,2})\delta^{(4)}(k_{c,A}+k_{c,B}-p_{c,1}-p_{c,2})\nonumber \\
 &  & \times\frac{1}{2}\int d^{2}\mathbf{b}_{c}\exp\left[i(\mathbf{k}_{c,A}^{\prime}-\mathbf{k}_{c,A})\cdot\mathbf{b}_{c}\right]\mathbf{b}_{c,j}[\Lambda^{-1}]_{\;j}^{\nu}\frac{\partial(\beta u_{\rho})}{\partial X^{\nu}}\nonumber \\
 &  & \times\left( p_{A}^{\rho}-p_{B}^{\rho}\right) f_{A}\left(X,p_{A}\right)f_{B}\left(X,p_{B}\right)\nonumber \\
 &  & \times\sum_{s_{A},s_{B},s_{1},s_{2} }\sum_{\text{color}}2 s_{2} \mathbf{n}_{c} \mathcal{M}\left(\{s_{A},k_{c,A};s_{B},k_{c,B}\}
 \rightarrow\{s_{1},p_{c,1};s_{2},p_{c,2}\}\right)\nonumber \\
 &  & \times\mathcal{M}^{*}\left(\{s_{A},k_{c,A}^{\prime};s_{B},k_{c,B}^{\prime}\}
 \rightarrow\{s_{1},p_{c,1};s_{2},p_{c,2}\}\right).
 \label{eq: N2}
\end{eqnarray}
Here the 3D form of the term $\epsilon^{0j\rho\nu}\partial(\beta u_{\rho})/\partial X^{\nu}\mathbf{e}_{j}=2\nabla_{X}\times(\beta\mathbf{u})$,
with $\mathbf{e}_{j}$ ($j=x,y,z$) being the basis vector in the
laboratory frame, $\beta\equiv1/T(x)$, $u^{\rho}$ is the fluid four-velocity,
$\mathbf{u}$ is the spatial part of $u^{\rho}$, $X$ and $y$ are defined as $X \equiv \frac{1}{2}(x_{A}+x_{B})$
and $y \equiv x_{A}-x_{B}$ for two incoming partons that are located at $x_{A}=(t_{A},\mathbf{x}_{A})$
and $x_{B}=(t_{B},\mathbf{x}_{B})$, $\mathbf{b}_{c}$ is the transverse part of $y_c=(0,\mathbf{b}_{c})$.
$\mathbf{n}_c=\hat{\mathbf{b}}_{c}\times\hat{\mathbf{p}}_{c,A}$ is the normal direction of the reaction plane of the scattering, $s_2=\pm 1/2$ denote the spin state of the quark or anti-quark along $\mathbf{n}_c$,
$v_{c,A}=|\mathbf{p}_{c,A}|/E_{c,A}$ and $v_{c,B}=-|\mathbf{p}_{c,B}|/E_{c,B}$
are the longitudinal velocities in the CMS (with $\mathbf{p}_{c,A}=-\mathbf{p}_{c,B}$),
$f_{A}$ and $f_{B}$ are the phase space Boltzmann distributions
for the incident particle $A$ and $B$ respectively.
The two Gaussian wave packets for the incoming particles are given by
\begin{equation}
\phi_{i}(\mathbf{k}_{i}-\mathbf{p}_{i})=\frac{(8\pi)^{3/4}}{\alpha_{i}^{3/2}}\exp\left[-\frac{(\mathbf{k}_{i}-\mathbf{p}_{i})^{2}}{\alpha_{i}^{2}}\right],\label{eq:wave-packet-gs}
\end{equation}
where $\alpha_{i=A,B}$ denote the width of the wave packet $A$ and $B$ respectively.
The definitions of other quantities can be found around Eqs. (18,21,22) of Ref. \cite{Zhang:2019xya}.
We see in Eq. (\ref{eq: N2}) that $\mathbf{P}$ is actually the difference between the number of quarks
(parton 2) with spin up and that with spin down at $X$.
The polarization rate at $X$ is given by
\begin{equation}
\frac{d\overline{\mathbf{P}}}{dt} = \frac{1}{n_q(X)}\sum_{A,B,1=\{q_{a},\bar{q}_{a},g\}}\frac{d^{4}\mathbf{P}_{AB\rightarrow1q}(X)}{dX^{4}} 
=\frac{1}{n_q(X)} \frac{\partial(\beta u_{\rho})}{\partial X^{\nu}}\mathbf{W}^{\rho\nu} , 
\end{equation}
where $n_q(X)$ denotes the quark number density at $X$ and is given by
\begin{equation}
n_q(X) = 6 \int \frac{d^3\mathbf{p}}{(2\pi )^3} \mathrm{exp} \left(-\beta \sqrt{m_q^2+\mathbf{p}^2}\right)
= \frac{3}{\pi ^2} m^2_q T K_2\left(\frac{m_q}{T} \right),
\end{equation}
where we have neglected the quark chemical potential and $K_2(z)$ is the modified Bessel function of the second kind. 
We note that using the thermal distribution for particle number (not for the spin) is just to estimate the polarization magnitude and compare with data. This does not contradict with the fact that our formalism does not assume thermal equilibrium for the spin degrees of freedom. Our numerical results show that the tensor $\mathbf{W}^{\rho\nu}$ has the form
\begin{equation}
\mathbf{W}^{\rho\nu}=W\epsilon^{0\rho\nu j}\mathbf{e}_{j} ,
\end{equation}
where we see that $\rho$ and $\nu$ should be spatial indices or $\mathbf{W}^{0\nu}=\mathbf{W}^{\rho0}=\mathbf{0}$.
Then Eq. (\ref{eq:polar-2}) is simplified as
\begin{equation}
\frac{d\overline{\mathbf{P}}}{dt} = 2\overline{W}\nabla_{X}\times(\beta\mathbf{u}),
\end{equation}
where $\overline{W}\equiv W/n_{\mathrm{q}}(X)$.

\begin{figure}
\begin{centering}
\includegraphics[scale=0.35]{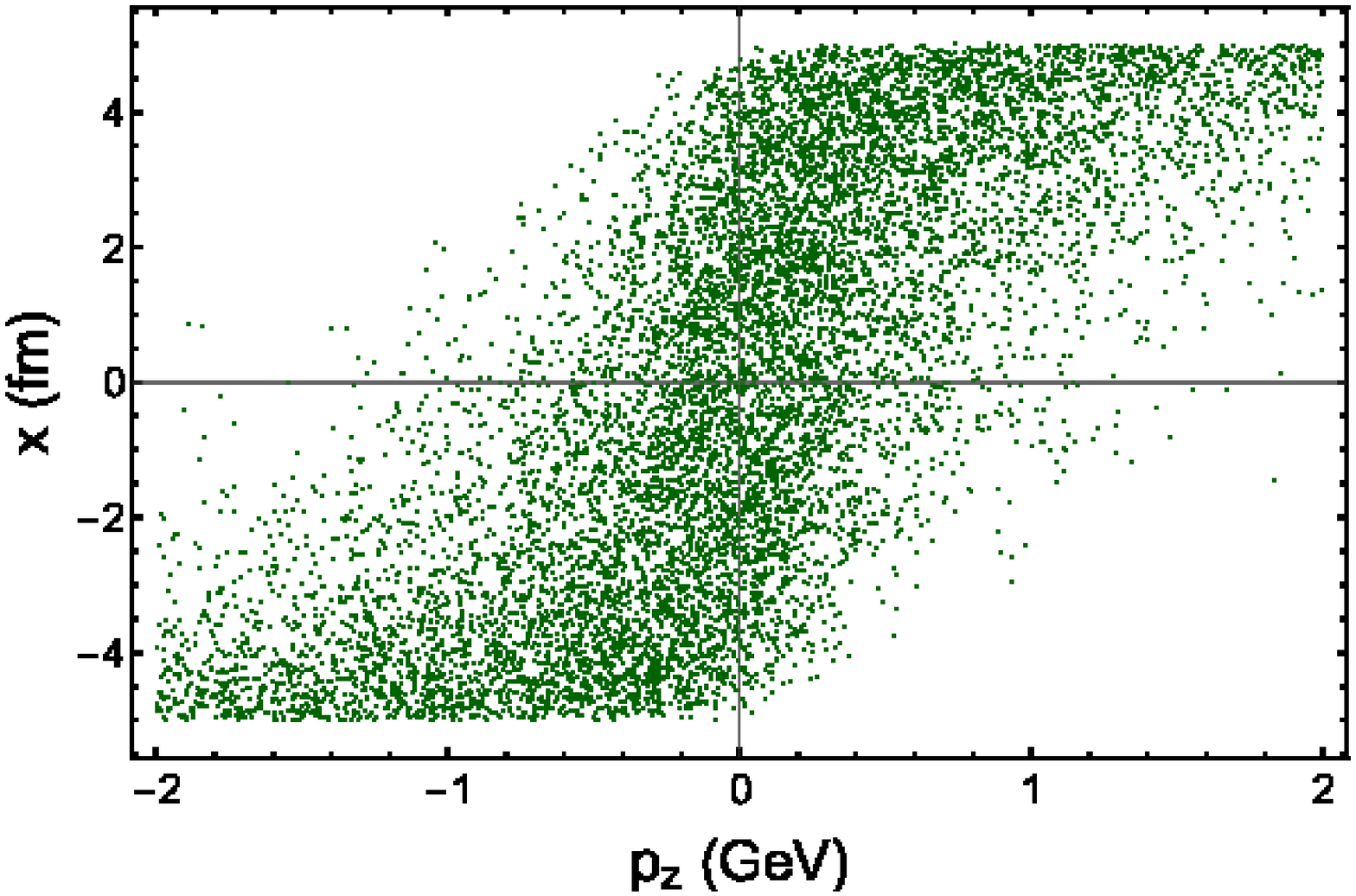}
\includegraphics[scale=0.35]{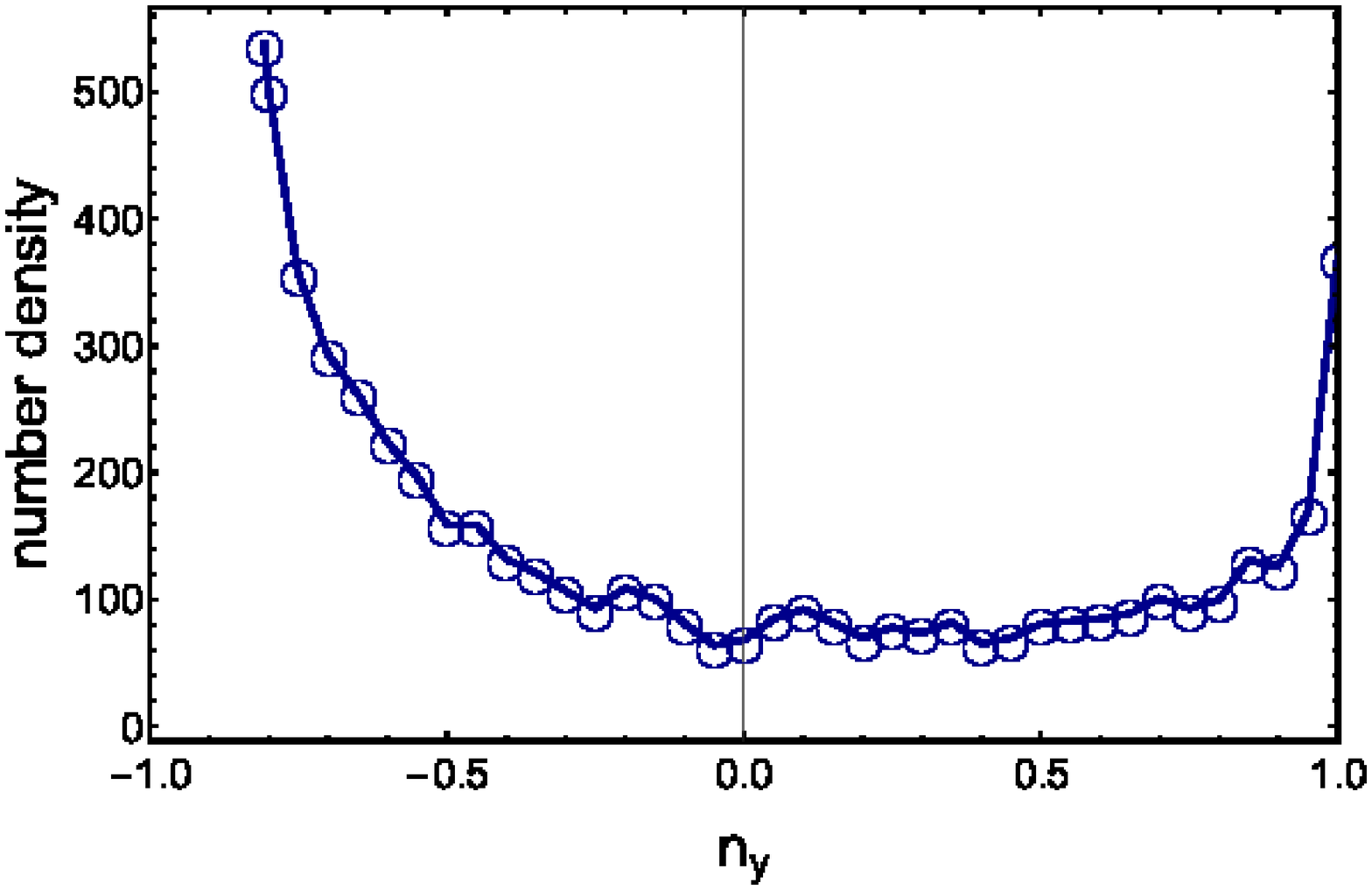}
\caption{Left panel: Distribution of $p_z$ in terms of $x$.
Right panel: Distributions of $n_{c,y}$, the $y$ component
of the orbital angular momentum direction. \label{fig:distributions-nxnynz}}
\end{centering}
\end{figure}

To illustrate the mechanism of the polarization effect in a fluid
moving along the $z$ direction, we propose a toy model in which particles
follow the Boltzmann distribution
\begin{equation}
f  =  e^{-\beta u\cdot p}=\exp\left[-\beta\gamma\left(\sqrt{m+\mathbf{p}^{2}}-axp_{z}\right)\right],\label{eq:dis}
\end{equation}
where $u(\mathbf{x})=\gamma(1,0,0,ax)$ denotes the fluid four-velocity
with $a$ being a small positive number, $\gamma=1/\sqrt{1-a^{2}x^{2}}$
is the Lorentz factor. Note that the velocity $u_{z}=\gamma ax$ depends
on $x$. We set the particle mass $m=0.2\text{GeV}$,
$a=0.2\;\text{fm}^{-1}$, and the phase space volume is a box of $\left[-5\text{fm},5\text{fm}\right]^{3}\times\left[-2\text{GeV},2\text{GeV}\right]^{3}$.
For the fluid velocity along the $z$ direction with an $x$-gradient,
the vorticity is in the $-y$ direction. We sample three-momenta of particles at each
space-time point. We randomly choose a pair of particles and transform
to their CMS frame according to their momenta (we use the index 'c' to indicate quantities in the CMS frame
of two particles). We limit
the time difference and longitudinal distance between two particles
to be small enough, i.e. $|\Delta t_{c}|<\Delta t_{\mathrm{cut}}$ and
$|\Delta z_{c}|<\Delta z_{\mathrm{cut}}$, where $\Delta t_{\mathrm{cut}}\sim\Delta z_{\mathrm{cut}}\sim 0$.
We also require that their distance in the transverse direction
be smaller than a cutoff, i.e. $|\Delta\mathbf{x}_{c,T}|<b_{0}$.
Then we can determine the direction of the orbital angular momentum of the pair
in the CMS frame, which we denote as $\mathbf{n}_{c}=(n_{c,x},n_{c,y},n_{c,z})$.
Figure \ref{fig:distributions-nxnynz} shows the distributions of
the $y$ component of $\mathbf{n}_{c}$. Note that only
the $n_{c,y}$ distribution is asymmetric with respect to negative and
positive $n_{c,y}$ while both $n_{c,x}$ and $n_{c,z}$ distributions are symmetric.
This gives rise to a negative $\left\langle n_{c,y}\right\rangle $,
which indicates that $\mathbf{n}_{c}$ favors the $-y$ direction
or the vorticity direction.

\section{Numerical results for polarization}
\begin{figure}
\begin{centering}
\includegraphics[scale=0.3]{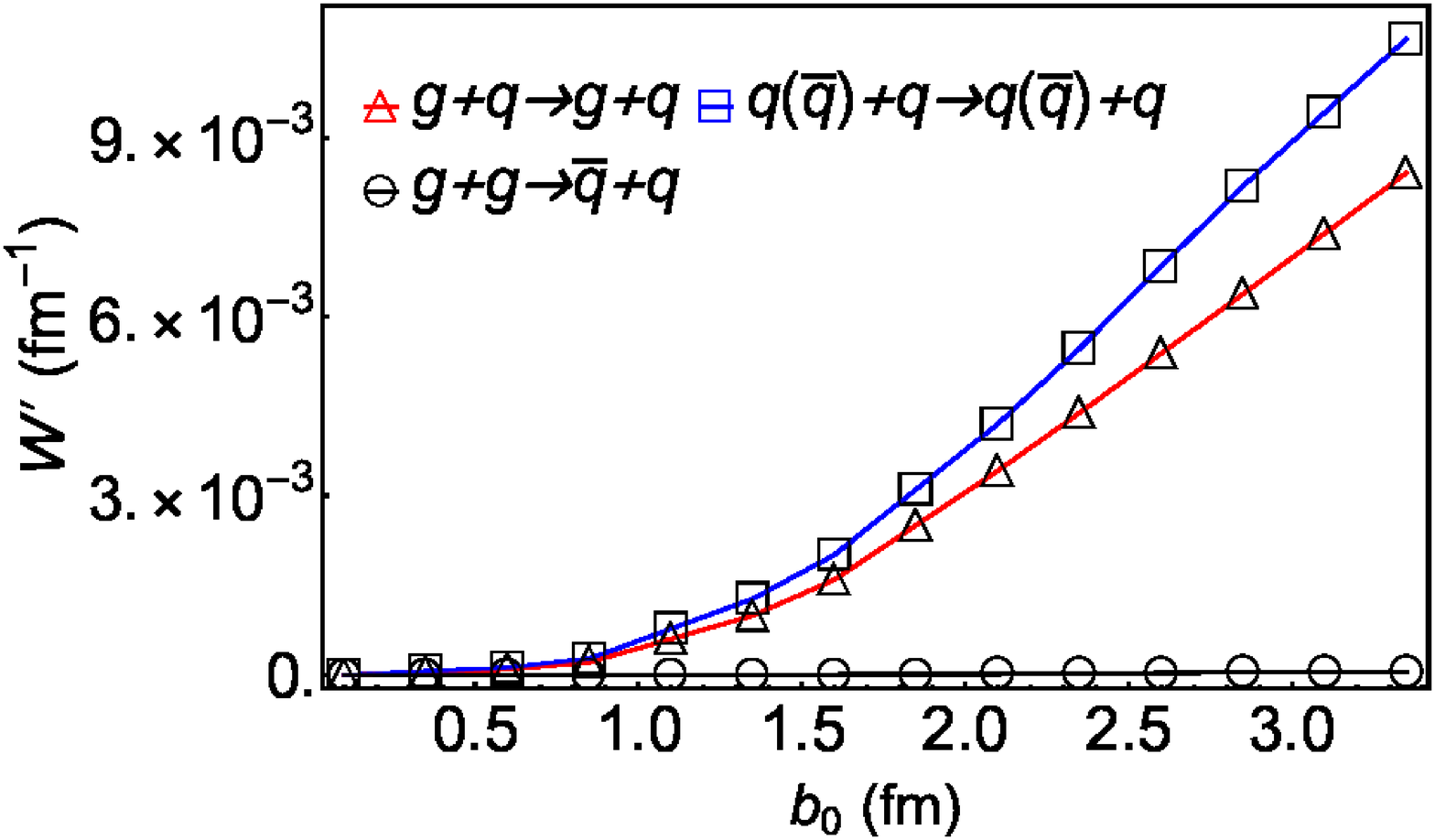}
\caption{$\overline{W}$ as a function of the cutoff $b_{0}$. \label{fig:-as-functions}}
\end{centering}
\end{figure}

In order to evaluate $\overline{W}$ numerically we need to set values of parameters:
the quark mass $m_{q}=0.2$ GeV for quarks of all flavors ($u,d,s,\bar{u},\bar{d},\bar{s}$),
the gluon mass $m_{g}=0$ for the external gluon, the internal gluon
mass (Debye screening mass) $m_{g}=m_{D}=0.2$ GeV in gluon propagators
in the t and u channel to regulate the possible divergence, the width of the Gaussian wave packet
$\alpha=0.28$ GeV, and the temperature $T=0.3$ GeV. 
The values of these parameters we choose are all generic ones 
for a parton system in high energy heavy ion collisions \cite{Zhang:2019uor}. 
Figure \ref{fig:-as-functions} shows the results of $\overline{W}$ as a function of the cutoff $b_{0}$ for the impact parameter, which corresponds to the coherence length of the colliding partons.

The STAR collaboration has measured the global polarization of $\Lambda$ and $\overline{\Lambda}$ in heavy ion collisons. We note that the polarization of $\Lambda$ and $\overline{\Lambda}$ come from $s$ and $\overline{s}$ respectively \cite{Yang:2017sdk}. We can estimate the polarization of quarks or anti-quarks (here we mean $s$ or $\overline{s}$) from $\overline{W}$ and compare with the STAR data at $\sqrt{s}=200$ GeV. We assume the average vorticity $(1/2)\langle \nabla_{X}\times(\beta\mathbf{u})\rangle _y \sim 0.04$, the time interval for the polarization is about 5 fm, $b_0=2.5$ fm, then we have $\overline{\mathbf{P}}_y \sim 0.3\% $, which agrees with the STAR data \cite{Adam:2018ivw}:  $0.277\pm 0.040 \pm 0.039\; [\% ]$ for $\Lambda $ and $0.240\pm 0.045\pm 0.061\; [\% ]$ for $\overline{\Lambda}$ .

\textbf{Acknowledgement.}
QW and RHF are supported in part by the National Natural Science Foundation
of China (NSFC) under Grant No. 11535012, 11890713 and 11847220.





\bibliographystyle{elsarticle-num}
\bibliography{ref-1}







\end{document}